# 概念层次的动态文本可视分析

向首兴[1)], 欧阳方昕[2)], 刘世霞[1)*]

[1)] (清华大学软件学院　北京　100084)
[2)] (微软(中国)有限公司　北京　100084)
(shixia@tsinghua.edu.cn)

**摘　要:** 分析社交媒体中关联主题在不同社会群体之间的流动模式有助于理解观点、信息和思想的传递. 已有的主题流动分析的工作大多是基于主题模型的, 只能通过查看包含该主题的文本来分析主题流动的原因. 这些文本数据量大且结构复杂, 难以分析. 为了解决这一问题, 使用概念对主题内部的内容进行概括, 提出了基于概念的动态文本可视分析方法, 用于展示主题内容的变化模式, 帮助分析主题流动的原因. 该方法使用流型线条展示概念流动模式, 并利用基于约束的 $t$-SNE 降维算法保证相邻时间段上概念投影分布的相似性, 从而保证流型线条的稳定性. 为了突出展示主题内概念的异常变化模式, 提出了一种异常检测技术用于定位概念剧烈变化的时间段并进行突出显示. 使用推特数据集进行定性评估和案例研究, 验证了所提出的可视分析方法的准确性和有效性.

# Concept-Based Visual Analysis of Dynamic Textual Data

Xiang Shouxing[1)], Ouyang Fangxin[2)], and Liu Shixia[1)*]

[1)] (*School of Software, Tsinghua University, Beijing*　100084)
[2)] (*Microsoft (China), Beijing*　100084)

**Abstract:** Analyzing how interrelated ideas flow within and between multiple social groups helps understand the propagation of information, ideas, and thoughts on social media. The existing dynamic text analysis work on idea flow analysis is mostly based on the topic model. Therefore, when analyzing the reasons behind the flow of ideas, people have to check the textual data of the ideas, which is annoying because of the huge amount and complex structures of these texts. To solve this problem, we propose a concept-based dynamic visual text analytics method, which illustrates how the content of the ideas change and helps users analyze the root cause of the idea flow. We use concepts to summarize the content of the ideas and show the flow of concepts with the flow lines. To ensure the stability of the flow lines, a constrained $t$-SNE projection algorithm is used to display the change of concepts over time and the correlation between them. In order to better convey the anomalous change of the concepts, we propose a method to detect the time periods with anomalous change of concepts based on anomaly detection and highlight them. A qualitative evaluation and a case study on real-world Twitter datasets demonstrate the correctness and effectiveness of our visual analytics method.

**Key words:** dynamic text visual analysis; dynamic textual data; concept analysis; anomaly detection

---

主题是文本中讨论的主要内容, 通常使用词频分布[1]来表示. 随着计算机网络和信息技术的广泛应用, 公众通过社交媒体参与到各种主题的产生和传播之中. 不同的主题拥有不同的生命周期. 有的主题生命周期长, 在多个社会群体之间流动; 有的主题生命周期短, 只存在于某些特定社会群体. 分析不同主题在不同社会群体之间的流动和传播, 可以帮助了解社会中各种主题的出现和发展、多个主题的收敛和发散, 以及各个社会群体在其中扮演的角色.

上述信息在市场分析和舆情分析等方面有着重要的作用. 例如, 对主题流动和传播的分析, 可以揭示不同群体的个体在多个社会问题上的领导力和影响力. 主题在这些个体之间的流动模式可以揭示政党在一系列相互关联的主题上的领先和滞后关系, 帮助政党及时回应竞争对手的声音, 借此更好地准备政治竞选. 领先-滞后关系指的是 $A$ 组个体(领先)所讨论的主题被 $B$ 组个体(滞后)跟随. 在真实需求的驱动下, 跟踪主题的流动并理解造成流动的原因, 分析群体内部和之间的领先-滞后关系也显得更加重要.

现有的主题分析方法多是基于主题模型的. 这些方法主要关注主题的流动, 可以方便地跟踪主题的流动和分析群体内部和之间的领先-滞后关系, 但是忽视了对主题内容的展现, 导致无法便捷地分析主题流动的原因. 具体来说, 在分析主题流动原因时, 仅仅查看主题中的关键词是无法深入了解主题内容的, 因此通常需要查看包含该主题的具体文档内容. 但这些文档数据往往数量很大, 可能涉及不同方面的内容, 并且这些内容会随时间而变化, 给用户的分析带来了 2 项技术挑战. 首先, 简单有效地展现主题内部内容的分布及其变化, 帮助用户探索主题流动的具体原因. 其次, 有效定位主题内容出现变动的时间段并突出展示.

为了解决上述挑战, 本文在 Wang 等[2]工作的基础上, 引入更细粒度的概念[3], 增加概念流的可视化部分, 以展示主题内部的概念, 帮助用户快速分析主题流动原因. 概念是与主题相关文本中的重要词汇. 为了更好地帮助用户理解概念的变化, 本文提出了概念流的可视化技术. 该技术使用流型线条表示主题中的概念, 清晰展示主题中包含的主要概念和这些概念分布随时间的变化. 概念流将每个时间点的概念投影在一维平面, 并将各个时间点的投影结果按顺序连接起来. 概念流中的一条流型线条代表一个概念随时间的变化模式. 为了同时保持概念分布的局部结构和概念流的连续性, 本文使用 Liu 等[4]提出的基于约束的 $t$-SNE 降维算法来进行投影. 结合概念流的应用场景, 使用上一个时间片段的投影结果作为下一个时间片段的约束, 从而保证相邻时间片段的分布稳定性和整个概念流的连贯性. 为了对主题内容进行恰当的时间切片, 突出展示主题内容的变化, 本文设计了基于异常检测的时间切片算法, 使用 $t$ 分布假设检验判断相邻时间片段的概念频率向量分布是否相同. 分布不同的相邻时间片段被认为是异常时间点, 异常时间点的概念分布会突出展示.

本文的主要工作是:

(1) 提出了一个可视分析系统, 帮助专家理解和分析一组相互关联的主题在不同社会群体内部和之间的流动以及变化的原因;

(2) 提出了一种基于异常检测的时间切片算法, 通过捕捉分布变化的时间片段, 展现主题内容分布变化的细节;

(3) 利用基于约束的 $t$-SNE 降维算法, 使得在保留概念分布局部结构的同时, 保证概念流的连贯性.

## 1 相关工作

本文工作与动态文本主题分析相关. 本节将动态文本主题分析方法分为单源动态文本和多源动态文本主题分析方法, 并从有效性和可扩展性对相关工作进行分析和探讨. 有效性包括信息准确性和信息丰富性, 其中信息准确性指的是信息符合文档中的主题分布并具有可用性; 信息丰富性则是指可以依赖该信息分析主题和它们之间的关联随时间的变化模式. 可扩展性指的是能够支持对大量主题进行分析.

### 1.1 单源动态主题分析的研究现状

单源动态主题分析指的是分析来自单个文本源, 带有时间序列关系的文本内容的主题. 单源动态主题分析方法可分为非层次化主题分析和层次化主题分析. 非层次化主题分析中, 主题之间没有层级差异, 即所有主题都在同一层级.

Blei 等[1]提出的动态主题模型(dynamic topic models, DTM), 通过控制每一时间段参数与上一时间段参数的差距, 得到随时间平滑变化的主题. 为了解决 DTM 信息丰富性的问题, Gao 等[5]利用增

量式层次化狄利克雷过程来提取主题的出现、流动和终止. Ahmed 等[6]提出一个统一的系统, 结合中餐馆过程(Chinese restaurant process, CRP)和隐含狄利克雷分布(latent Dirichlet allocation, LDA)模型, 与此同时提取文本中的人物、时间、地点和主题. 结合文本之间的引用关系, Wang 等[7]提出 Citation-LDA 模型, 提取主题内容、强度、重要性和主题间依赖关系, 以及它们随着时间的变化. 如何展示挖掘到的主题信息也是研究者关注的问题. Havre 等[8]的工作 ThemeRiver 中最先使用河流的视觉隐喻来表示主题的流动, 使用条带表示主题, 条带宽度表示主题热度. 为展示更多主题信息, TIARA[9-12]将主题的关键词布局在条带上; Visual Backchannel[13]在河流形式上, 增加了螺线图展示主题对应的人; ParallelTopics[14]增加了平行坐标以展现文档中主题的概率分布. 为了分析主题间的相互关联, TextFlow[3]使用桑基图(Sankey diagram)表示主题的分裂和合并. 非层次化方法的有效性较好, 但可扩展性较差, 难以处理大数量的主题.

层次化方法不仅挖掘主题的信息, 同时也利用树的结构来组织得到的主题, 因此具有较好的可扩展性. 挖掘层次化主题的方法主要是带约束的层次化聚类, 使用 $t$ 时间段的结果作为约束来得到 $t$+1 时间段的结果. 约束可分为成对约束和三元组约束 2 类. 成对约束是约束 2 个样本在或者不在同一个类里. Davidson[15]在 2009 年指出, 若将成对约束作为硬约束, 即必须满足的约束, 那么凝聚式层次聚类方法可能没有可行解. Miyamoto 等[16]提出使用软约束的方法来避免无可行解的问题. 使用成对约束能够生成比无约束下更好的聚类结果, 但因为成对约束不包含层次信息, 因此无法刻画层次化主题的结构. 三元组约束是指2个样本要在与第 3 个样本合并前先合并, 包含了层次的信息. 现有的利用三元组约束的方法可大致分为2类, 基于度量的方法和基于实例的方法. 基于度量的方法是通过三元组约束得到样本的相似矩阵, 进而应用到算法中; 而基于实例的方法则是在自底向上的聚类过程中要求结果满足约束. 使用三元组约束可以部分刻画层次, 因此聚类效果优于成对约束, 但仍有不足, 仅能刻画二叉树的结构, 不能刻画多叉树结构. 已有的单源动态文本的层次化主题可视分析有 ParallelTopics[14]和 FluxFlow[17]等. 但这些方法中不同时间段的主题树之间是没有关联的, 因此对主题的分裂和合并展示不足.

### 1.2　多源动态主题分析的研究现状

多源动态主题分析指的是分析多个文本源的时间序列文本的主题. 分析的主题信息可分为不同文本源间的共有和独有主题, 不同文本源间的全局与局部领先-滞后关系. 早期的挖掘共有和独有主题的工作[18]使用统一的主题模型对不同文本源进行建模, 忽略了不同文本源主题上的差异, 提取的文本准确性不佳. 为解决该问题, 演化层次狄利克雷过程(evolutionary hierarchical Dirichlet processes, EvoHDP)[19]在提取多元静态文本共有主题的层次化狄利克雷过程(hierarchical Dirichlet processes, HDP)上添加了时间维度的依赖, 以提取动态文本的共有主题. 为了更好地挖掘独有主题, Hong 等[20]提出了具有时间依赖性的主题模型. 以上方法能较好地挖掘共有和独有主题, 但未考虑相关主题间的领先-滞后关系, 在信息丰富性上有所欠缺. 因此研究者提出了挖掘主题全局领先-滞后关系的方法, 大致可分为基于短语的方法、基于文档的方法和基于主题的方法. 基于短语的方法[21-23]中, 通过平移不同文本源中短语的时间序列来进行匹配, 匹配程度最高的平移时间即为文本源间的全局领先-滞后时间. 基于文档的方法[24-26]则是通过概率模型来检查文档间是否有领先-滞后关系. 而基于主题的方法[27-28]是使用 LDA 主题模型来提取主题间的领先-滞后关系. 全局领先-滞后关系能够刻画文档的领先程度, 衡量文档的影响力, 但无法刻画领先-滞后关系随时间的变化过程, 因此研究者提出了挖掘每个时间段中的局部领先-滞后关系的方法. TextPoineer[29]中, 若主题 $A$ 与上一个时间段中主题 $B$ 的相似度高于主题 $A$ 与下一个时间段中主题 $B$ 的相似度, 则在当前时间段中, $A$ 领先于 $B$. 该方法仅能分析同一主题在不同文本源中的领先-滞后关系, 不能应对多个相互关联的主题的情况. 而 Zhong 等[30]提出的通过词汇的时间序列的协整(cointegration)关系来决定领先-滞后关系的方法, 仅能处理多个互相关联的主题在2个文本源中的领先-滞后关系, 不能应对多于 3 个文本源的情况. 为了解决这一问题, IdeaFlow[31]使用随机游走相关模型来结合文本内容、词汇时间序列的协整关系和文本源之间的引用关系, 提高了挖掘模型的准确性; 利用张量统一的原理来结合多个文本源, 以分析多个相互关联的主题在多于3个文本源中的领先-滞后关系. IdeaFlow 能够帮助用户快速分析多个相互关联的主题在多个文本源中

的不同时间点的领先-滞后关系，但是在分析主题流动的原因时，需要查看与主题相关的文本，这些文本数据量大且结构复杂，难以分析．为了解决这一问题，本文在 IdeaFlow 的基础上，引入了更细粒度——概念，通过展示主题内部的概念，帮助用户快速分析主题流动的原因．具体来说，本文以螺纹线形式展现主题内部的概念，为了突出显示重要的概念，设计了一个异常检测技术来找到概念剧烈变化的时间，利用一种基于约束的 *t*-SNE 降维算法来展示概念随时间变化的模式以及概念间的相关关系．

## 2 需求分析与系统概览

本节介绍相关需求的收集以及基于需求所设计的可视分析方法．

### 2.1 需求分析

本工作的思想源于 TextFlow[3]．Cui 等[3]认为，尽管主题挖掘方法能够很好地挖掘出文本中的主题及其演变过程，但结果非常抽象，难以被用户理解．因此，TextFlow 中通过展示 3 个层次粒度的结果帮助用户理解主题的演变．而在分析多个相互关联的主题在多个动态文本源中的流动时，同样需要分析和理解主题流动的原因，因此需要概念粒度的主题内容展示．

为验证概念粒度的主题内容展示是否必要，本文邀请一位媒体传播领域专家(*E*)进行采访，该专家想要了解政党和政治人物在社交媒体上的交际策略，希望可视分析系统能帮助他分析各种社会群体在社会问题中扮演的角色．本文利用 IdeaFlow 系统找出一些分析主题的案例，并向专家演示．专家看完演示后，使用 IdeaFlow 系统对他自己的数据集进行了分析．该过程耗时约 2 h．最后，专家对 IdeaFlow 的优缺点进行评价:

(1) 探索主题流动的原因能有效地帮助分析者判断观察到的主题流动是否真实;

(2) 概念粒度的主题内容展示能帮助分析者分析主题流动的细节和原因;

(3) 如果发现观察到的主题流动是错误的，对提取主题流动的方法进行针对性地调整是非常有用的．

专家的评价验证了需求：概念粒度的主题内容展示是有价值的．接着，本文设计概念粒度展示主题内容的可视化原型，并定期与专家讨论，迭代式地完善原型．基于与专家整个过程的讨论和反馈，得到以下 2 个需求:

**R1**．以简单易懂的形式展现主题内容的变化，特别是内容分布的局部结构，并且时间段之间的分布应该具有连贯性;

**R2**．定位主题内容发生变动的时刻，从而展现能够帮助专家判断的有价值的主题内容．

### 2.2 系统概览

基于以上需求，本文在 IdeaFlow[31]基础上开发了新的系统，它由图 1 所示数据处理和可视化 2 个模块组成．给定一组推特、用户和用户关注关系等信息，数据处理模块从中提取主题、主题流动和主题内部的概念，并进行切片和投影得到概念的流动信息，可视化模块展示提取出的信息，以供用户交互式探索和分析．数据处理模块包含 3 个方法：基于随机游走的相关性模型用于提取主题、主题的流动及主题内部的概念；基于异常检测的时间切片算法用于定位主题内容分布出现突变的时间段，进而突出展示；基于约束的 *t*-SNE 用于投影主题内部的概念，以流型线条的形式简单易懂地展示概念的流动，帮助专家分析主题流动的原因．可视化模块包括主题流图和概念流图：主题流图用以描述相关主题在单个文本源中和多文本源间的流动；概念流图用以展现主题内部概念的分布和流动(R1, R2)．该系统还提供多种交互探索功能，以帮助专家探索和比较详细信息．例如，图 2 中 D 区域以气泡树的形式展示了主题及其层次结构；图 2 中 E 区域展示了当前主题下领先和滞后的关键词、用户和推特，其中关键词和用户都是以词云的形式展示．

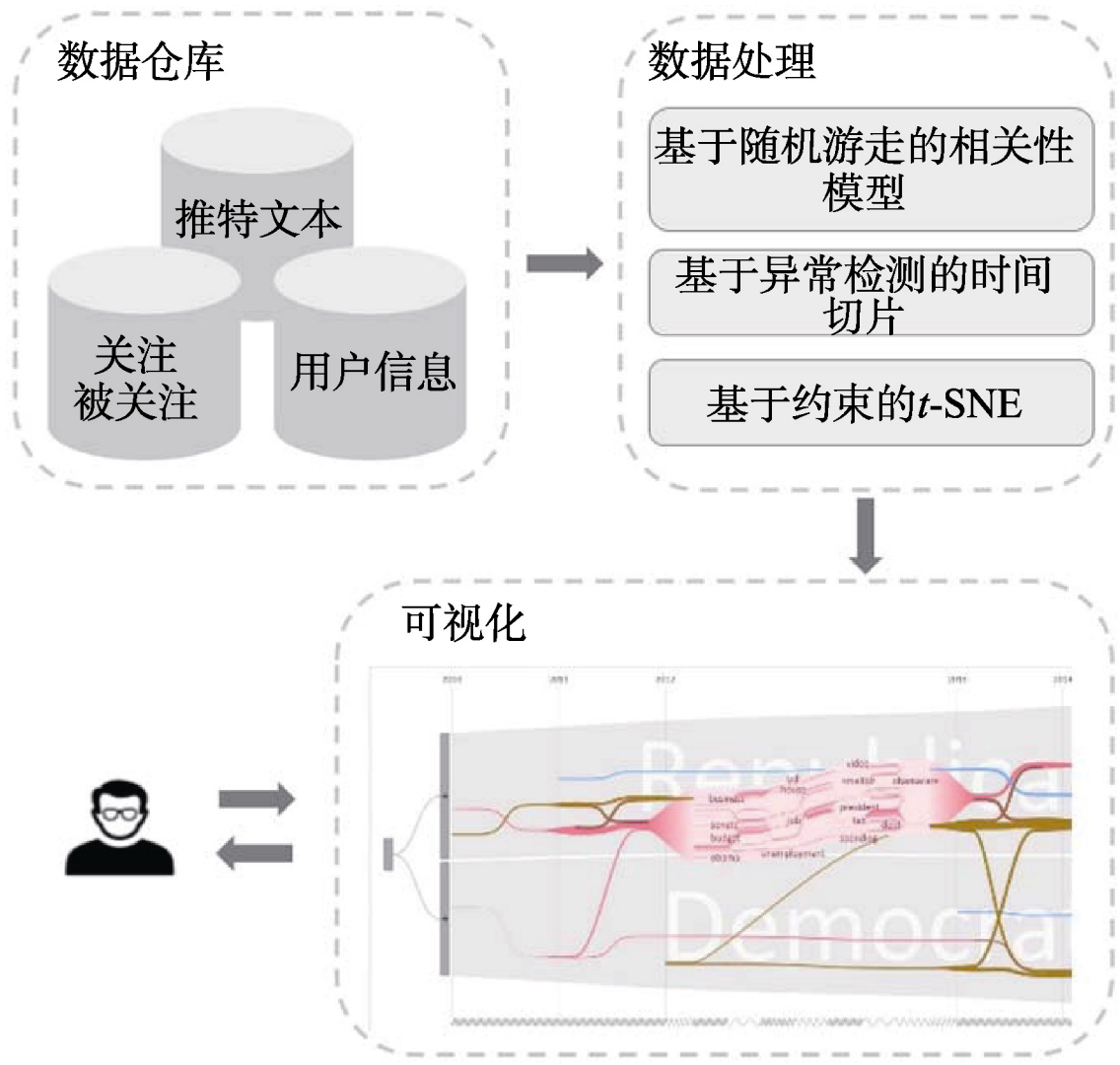

图 1 系统概览图

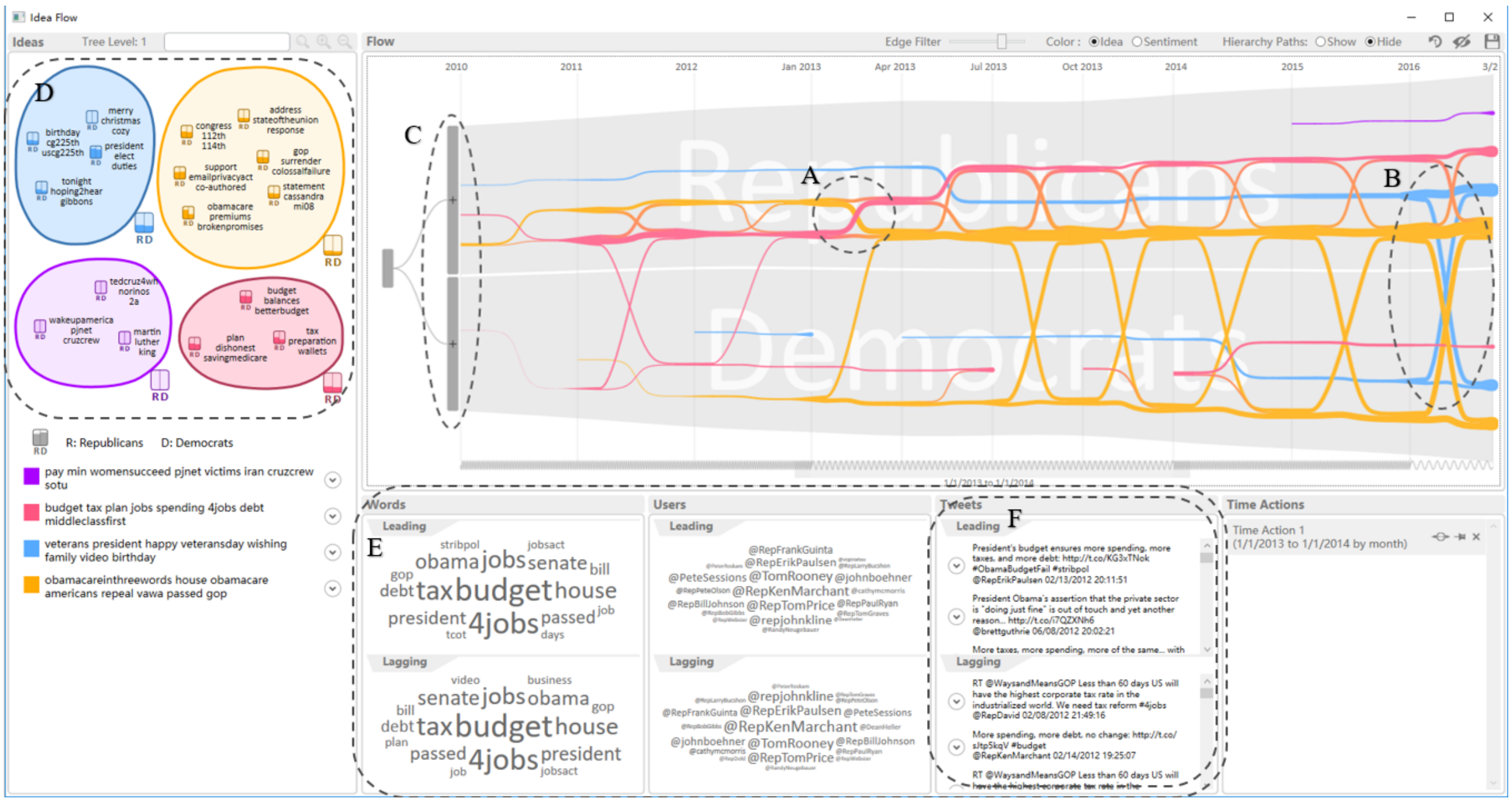

图 2　加载美国政党数据集的系统界面

## 3　概念流

为了满足 R1，本文设计了一种如图 3 所示可视化形式——概念流，能够清晰地展示主题中包含的主要概念和这些概念分布随时间的变化．受 Ferstl 等[32]的工作启发，本文使用流型线条来表示主题中的概念，以流型线条的合并和分裂来表示概念之间关联关系的变动．图 3 是概念流的表示示例，每个条带代表一个主题．从图 3 中 A 区域可以看到主题内部的概念，每一线条代表一个概念，线条的宽度代表概念的重要程度；其中，横轴是时间轴，线条高度的变化表示概念随时间的变化．在每一个时间切片中，线条片段的高度是由概念的特征向量投影得到的．具体来说，通过以下几个步骤获得主题内部概念对应的流型线条：

Step1．提取概念的特征向量以描述概念．

Step2．根据时间进行切片，突出展现主题内容发生变动的时刻(R2)．

Step3．对概念特征向量进行降维，将概念投影到一维空间，从而得到每个时间段的一维概念分布．并且相邻时间段的概念分布之间应该具有连贯性．

### 3.1　数据处理

本文从主题相关的文本中提取词频高的 top-$K$ 的关键词作为主题中的概念．常用的单词的特征向量是词嵌入向量，可以表示单词之间的语义关系；然而，词嵌入向量中不包含该单词所属文本的信息和发布该概念的用户的信息．而本文中推特间的用户关系是很重要的，因此仅仅使用词嵌入

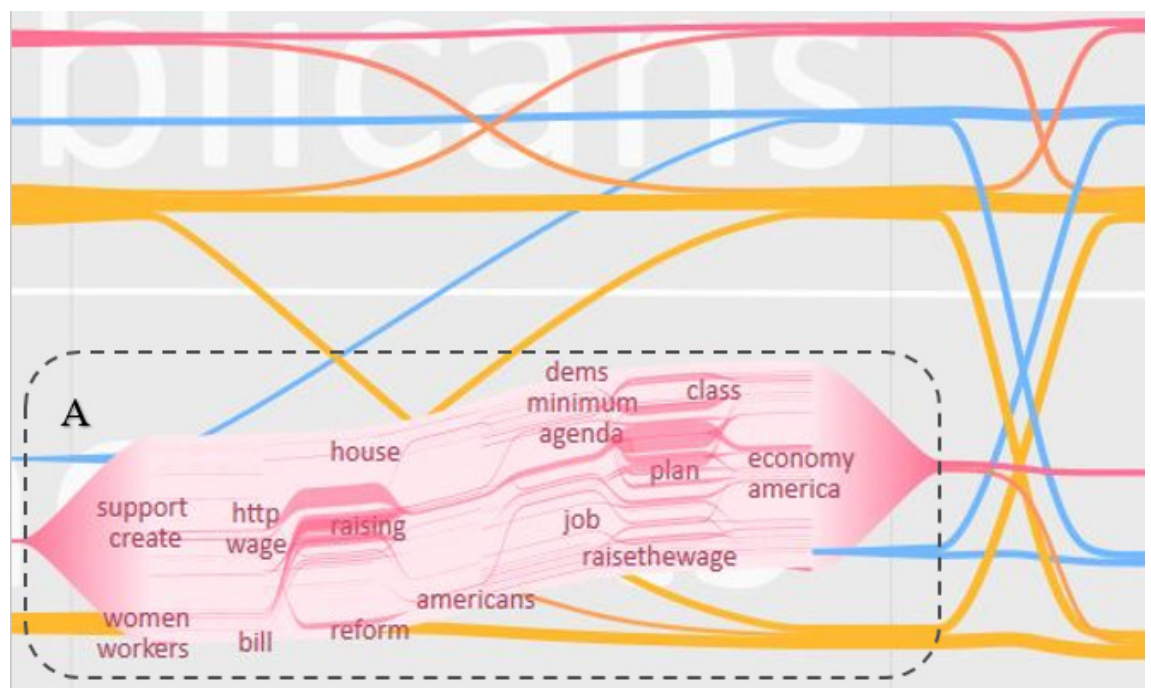

图 3　概念流

向量不能很好地针对本文的应用场景．本文结合词嵌入向量 $\boldsymbol{D}$ 和文本信息向量 $\boldsymbol{F}$ 来提取概念的特征向量．具体来说，参照信息检索领域提取文本特征向量的方式，假设与主题 $A$ 相关的文本中共包含 $M$ 个概念，在进行时间切片后，当前时间段的文本数量为 $N_t$，那么这些概念的文本信息向量矩阵即为 $\boldsymbol{F}\in\mathbb{R}^{M\times N_t}$．若第 $m$ 个概念出现在第 $n_t$ 条文本中，则 $\boldsymbol{F}_{mn_t}=1$；否则 $\boldsymbol{F}_{mn_t}=0$．通过连接词嵌入向量和文本信息向量得到概念的特征向量 $\boldsymbol{G}=[\boldsymbol{D},\alpha\boldsymbol{F}]$．另外，实验表明，某些概念之间的语义相关性并不高，但它们代表了事件中的一些关键信息，能够帮助用户发现和分析突发事件．因此，文本信息向量应当有较高的权值，本文中采用 $\alpha=0.9$．

### 3.2　可视化方法

#### 3.2.1　基于异常检测的时间切片算法

图 3 中的每一个条带都代表一个主题，包含了

具有较长时间跨度的文本信息. 为了展示主题中概念的变化过程, 本文采用了时间切片的方法. 将这段较长的时间切分成多个时间片段, 计算每个时间片段主题中概念的分布, 并同时展现多个时间段的概念分布, 以展示出主题中概念随时间变化的过程. 具体地讲, 以图 3 中 A 区域的形式来展现主题内部的概念, 其中每一个线条都代表一个概念, 当线条出现较大的偏移时, 就说明出现了突发事件. 通过分析突发事件, 用户就能探究主题流动的原因. 但发生突发事件的时间仅占整个时间段的很小部分, 简单的均匀切片不能很好地展现突发事件. 基于通过恰当地不均匀切片, 因此本文参照 Lu 等[33]的方法, 对词频向量的分布进行异常检测, 以此进行不均匀切片. 具体地说, 先使用较小的粒度进行均匀时间切片, 接着使用 $t$ 分布来衡量相邻 2 个时间段的词频向量的相似程度, 当相邻时间段的词频向量不相似时, 则以此异常时间点进行切分, 就能得到一系列不均匀的时间段, 这样的时间切片算法能够较好地展现突发事件. 另外, 每一个条带的时间跨度是一年, 而突发事件的时间跨度不超过 20 天, 因此本文参照 Wang 等[31]的方法, 使用图 4 所示弹簧的形式来表示时间轴, 弹簧的紧密程度表示时间跨度的大小.

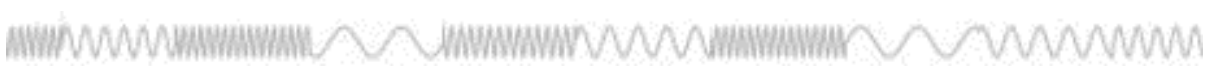

图 4 弹簧时间轴表示图

3.2.2 基于约束的 $t$-SNE 降维算法

为了满足 R1, 本文采用流型线条的形式来展示概念, 因此需要将概念投影到一维空间, 并且相邻时间段的投影分布之间应该相似, 以保证概念流的连贯性. 因为需要通过观察概念聚类的现象来分析主题内容的变化, 而投影算法应该能够较好地保持分布的局部结构, 所以本文选择了可视化领域最为经典的降维算法 $t$-SNE; 但 $t$-SNE 投影结果的全局结构是不稳定的, 相似的高维数据可能得到具有不同全局结构的投影结果; 因此, 本文采用了基于约束的 $t$-SNE 算法[4]. 在 Liu 等[4]的工作中, 通过该算法控制投影结果中各类别集群的全局分布, 本文的应用场景中可使用该算法来控制相邻时间段的概念投影结果的分布相似性. 具体地说, 在 $t$ 时间段采用 $t-1$ 时间段的投影结果作为约束来进行投影, 那么当概念分布未产生突变时, $t$ 时间段的投影就与 $t-1$ 时刻的投影相似, 保证了概念流的连贯性. 下面简单介绍一下基于约束的 $t$-SNE 算法. 原始的 $t$-SNE 算法中, 首先计算点在高维空间中的联合概率分布矩阵 $\boldsymbol{P}$ 和点在投影后空间中的联合概率分布 $\boldsymbol{Q}$, 即

$$p_{ij}=\frac{\exp(-\|\boldsymbol{x}_i-\boldsymbol{x}_j\|^2/2\sigma^2)}{\sum_{k\neq l}\exp(-\|\boldsymbol{x}_k-\boldsymbol{x}_l\|^2/2\sigma^2)},$$

$$q_{ij}=\frac{(1+\|\boldsymbol{y}_i-\boldsymbol{y}_j\|^2)^{-1}}{\sum_{k\neq l}(1+\|\boldsymbol{y}_k-\boldsymbol{y}_l\|^2)^{-1}}.$$

其中, $p_{ij}$ 和 $q_{ij}$ 分别指的是在高维空间和投影空间任意选择 2 个点时选到点 $i$ 和点 $j$ 的概率; $\boldsymbol{x}_i$ 和 $\boldsymbol{y}_i$ 分别指的是点 $i$ 在高维空间和投影空间的坐标. 原始的算法中, 通过 KL 散度来衡量 2 个分布间的距离, 因此得到代价函数

$$C=\mathrm{KL}(\boldsymbol{P}\|\boldsymbol{Q})=\sum_i\sum_{j\neq i}p_{ij}\log\frac{p_{ij}}{q_{ij}} \tag{1}$$

由式(1)所示代价函数, 投影问题可以使用梯度下降的方法来求解. 基于约束的 $t$-SNE 在原始算法的基础上加上了用户约束的考量, 其定义的代价函数为 $f_{\mathrm{cost}}=\alpha\mathrm{KL}(\boldsymbol{P}\|\boldsymbol{Q})+(1-\alpha)\mathrm{KL}(\boldsymbol{P}_{\mathrm{c}}\|\boldsymbol{Q}_{\mathrm{c}})$. 其中, 第 1 项 KL 散度项与原始的 $t$-SNE 相同, 用来保持投影结果的准确性; 第 2 项 KL 散度项则是保持投影结果的稳定性; $\alpha\in[0,1]$, 用来控制 2 项 KL 项的权重. 假设共有 $m$ 个概念, 对每一个概念引入一个虚拟点作为约束, 其高维空间的坐标与其对应的概念相同, 其投影空间的坐标是其对应的概念在上一时间段的投影坐标. 那么 $\boldsymbol{P}_{\mathrm{c}},\boldsymbol{Q}_{\mathrm{c}}\in\mathbb{R}^{m\times 1}$ 分别表示概念与其对应虚拟点在高维空间和投影空间的相似度, 若第 $i$ 个概念在当前时间段发生了突变, 则 $(\boldsymbol{P}_{\mathrm{c}})_i=1$; 否则 $(\boldsymbol{P}_{\mathrm{c}})_i=0.5$. 而 $(\boldsymbol{Q}_{\mathrm{c}})_i=\left(1+\|\boldsymbol{y}_i-\boldsymbol{y}'_i\|^2\right)^{-1}$, 表示的是当前投影坐标 $\boldsymbol{y}_i$ 与上一时间段投影坐标 $\boldsymbol{y}'_i$ 的距离. 因此, 通过第 2 项 $\boldsymbol{P}_{\mathrm{c}}$ 和 $\boldsymbol{Q}_{\mathrm{c}}$ 的 KL 散度, 可以保证相邻时间段的概念投影分布的相似性.

3.2.3 流型线条可视化

本文使用流型线条来表示概念的流动. 首先, 从主题相关的文本中提取出频率最高的 $k$ 个关键词作为概念; 接着使用基于异常检测的时间切片算法将主题相关的文本按照时间进行切分; 然后对于每一个切片后的时间片段, 提取每个概念的特征向量, 并使用基于约束的 $t$-SNE 将概念投影到

一维空间; 最后, 将概念的一维空间投影按照时间段的顺序从左到右堆叠, 连接相邻时间段里的相同概念, 则每个概念展示为一个流型线条. 但仅仅通过图 3 所示线条表示概念流动比较抽象, 用户无法辨别概念的内容, 因此本文利用一个基于网格搜索的标签布局算法, 将概念对应的关键词显示在对应的线条周围. 当专家将鼠标悬停在关键词上时, 关键词标签和对应的线条会高亮; 若专家点击关键词, 则会在信息面板显示包含对应概念的具体文本内容. 结合展现的文本, 实验发现使用基于约束的 $t$-SNE 投影的结果聚类效果不佳, 因此展现的关键词较多, 其中包含某些不够重要的, 影响专家分析. 为解决这一问题, 本文使用 Story-Flow[34]中提出来的流线压缩算法, 以此缩小类内距离, 增大类间距离.

#### 3.2.4 交互探索环境

本文开发了一个可视分析系统来帮助专家理解和分析主题在不同社会群体内部和之间的流动以及变化的原因. 系统包含图 2 所示多个视图, 帮助专家从多个角度挖掘主题流动的信息. 图 2 中 D 区域所示为系统的概览视图, 是一棵表示主题结构的气泡树, 每一个气泡代表一个主题, 气泡大小编码主题的重要性, 通过气泡的包含关系来表示主题的层次结构. 图 2 中 A~C 区域都是系统的流视图的组成部分, 其中图 2 中 A 区域代表的是 2 个不同的文本源, 在图 2 所示的数据中表现为民主党和共和党; 图 2 中 A 和 B 区域则分别表示同文本源的主题流动关系和不同文本源之间的主题流动关系, 其中每一个颜色的条带代表一种主题, 条带的宽度编码的是主题的活跃程度, 条带的相交就代表主题的流动. 图 2 中 E 和 F 区域则是系统的信息视图, 图 2 中 E 区域展现的是文本中重要关键词的领先滞后关系, 帮助专家在关键词上对主题流动情况进行分析; 而图 2 中 F 区域展示的则是具体的文本内容, 例如, 推特数据集下的每条推特的文本内容. 除了图中展现的视图外, 点击流视图中的条带, 例如, 图 2 中 A 区域的条带, 就能对代表主题的条带进行展开, 进入到概念流视图观察和分析主题内部的概念流动情况.

## 4 实验与分析

为验证所提出可视分析方法的准确性和有效性, 本文对提出或使用的算法进行了定性实验, 并邀请专家 $E$ 在真实数据集上进行案例分析. 本节中介绍了定性实验过程和分析结果以及案例分析的内容和结论. 网络文本中包含大量的有价值的信息, 帮助分析网络文本具有极大的实用价值[35]. 本文使用的数据集是美国政党的推特数据集, 其中包含第 114 届美国议会成员的 1 102 个推特账号在 2010 年 1 月—2016 年 3 月发布的共 1 605 361 条消息, 后文中提到的推特数据集均指该数据集.

### 4.1 算法有效性验证

#### 4.1.1 时间切片算法有效性验证

为验证基于异常检测的时间切片算法的有效性, 本文在推特数据集上进行了实验. 基于异常检测的时间切片和均匀时间切片的对比结果如图 5 所示. 专家 $E$ 通过查看该处推特文本内容以及推特之间的转发关系和用户之间的关注关系, 发现其包含的概念中存在 2 个异常(概念分布变化较大的地方): 异常 1(图 5 中 A 和 C 区域)和异常 2(图 5 中 B 和 D 区域). 异常 1 中, 推特用户首先抱怨参议院近 4 年没有公布财政, 随后开始呼吁参议院通过“No Budget No Pay”提议, 即参议院如果不公布财政预算, 参议院人员将无法得到工资, 最终参议院迫于压力公布了财政预算. 从基于异常检测的时间切片算法的结果图(图 5 中 C 区域)中可以看到, 异常 1 中的相关概念被正确检测出(如“senate”, “budget”, “passed”等). 但是从均匀时间切片的结果图(图 5 中 A 区域)中无法看到这些相关概念. 这说明相比于均匀时间切片, 基于异常检测的时间切片算法可以更好地检测到与异常相关的概念. 异常 2 中, 推特用户开始讨论奥巴马医疗法案导致许多美国人民失去医疗保险以及美国政府面临关门等问题. 随后推特用户开始发起投票, 要求停止奥巴马医疗法案(“keep my health”). 从基于异常检测的时间切片算法的结果图(图 5 中 D 区域)中可以看到明显的与异常相关的概念: “obamacare”, “government”和“health”等. 虽然均匀时间切片的结果图(图 5 中 A 区域)中存在这些概念, 但是其不明显, 并且存在许多无关概念(“budget”和“loan”等). 这说明基于异常检测的时间切片算法能够突出与异常相关的概念, 并过滤掉与异常无关的概念.

由上述 2 个实验结果可以看出, 通过基于异常检测的时间切片, 能够更好地检测到与异常相关的概念, 过滤与异常无关的概念, 从而帮助专家 $E$ 分析主题流动的原因.

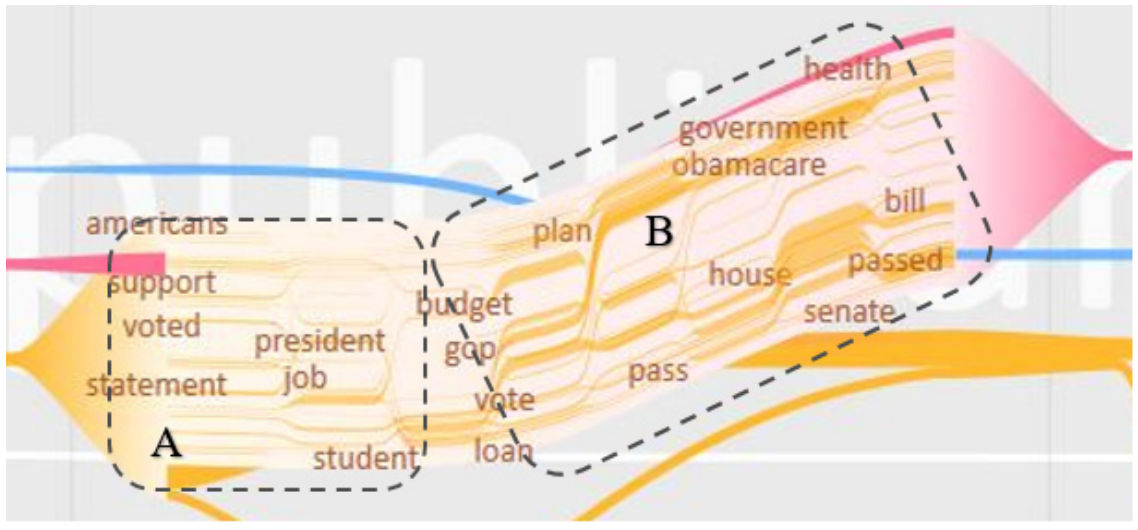

a. 均匀切片算法效果图

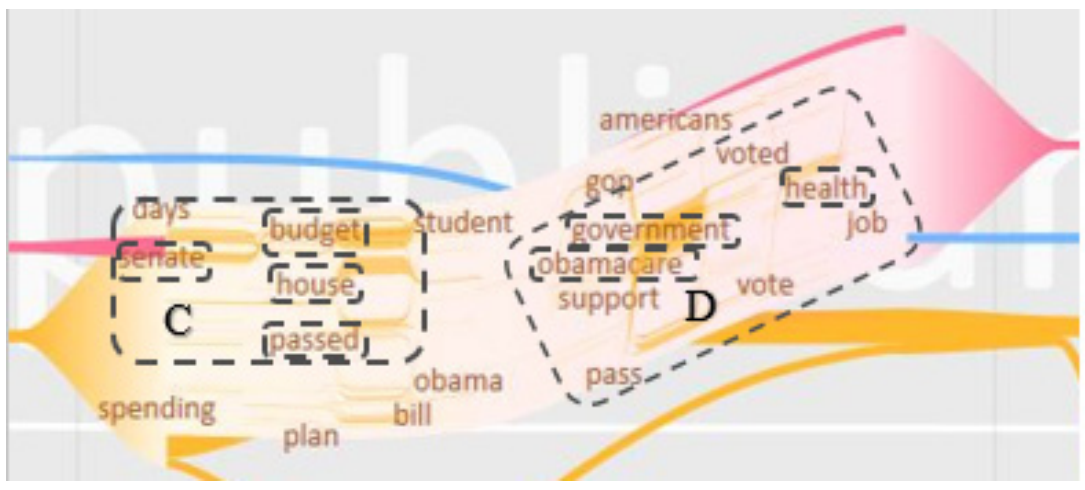

b. 基于异常检测算法效果图

图 5 基于异常检测的时间切片算法效果图

#### 4.1.2 基于约束的 *t*-SNE

为验证基于约束的 *t*-SNE 能够帮助用户更好地查看概念的变化, 进而分析主题流动的原因, 本文在推特数据集上进行了图 6 所示实验. 专家可以从图 6 中 C 和 D 区域观察到“budget”等明显的概念聚类, 进而猜测主题流动的原因. 而在图 6 中 A 区域, 相邻时间段的概念分布差距很大, 专家很难观察到“budget”等重要概念聚类; 图 6 中 B 区域也有同样的问题, 因为分布的不集中, 重要概念聚类“debt”的文本标签未显示. 图 6 的对比结果说明, 使用基于约束的 *t*-SNE 来保证相邻时间片段的概念分布稳定性以及概念流的连贯性能帮助专家分析主题流动原因.

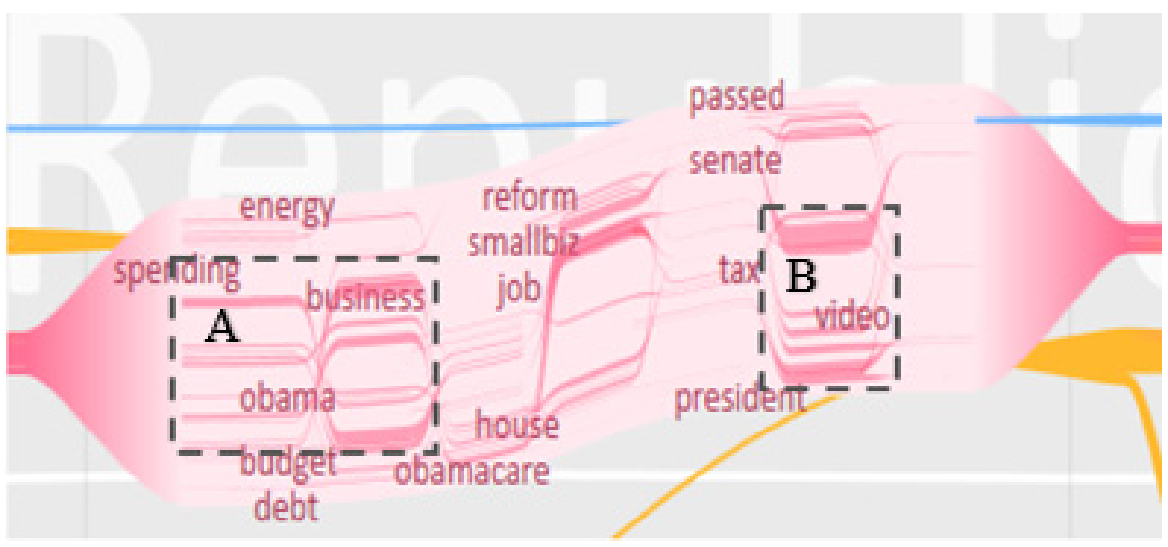

a. 无约束算法效果图

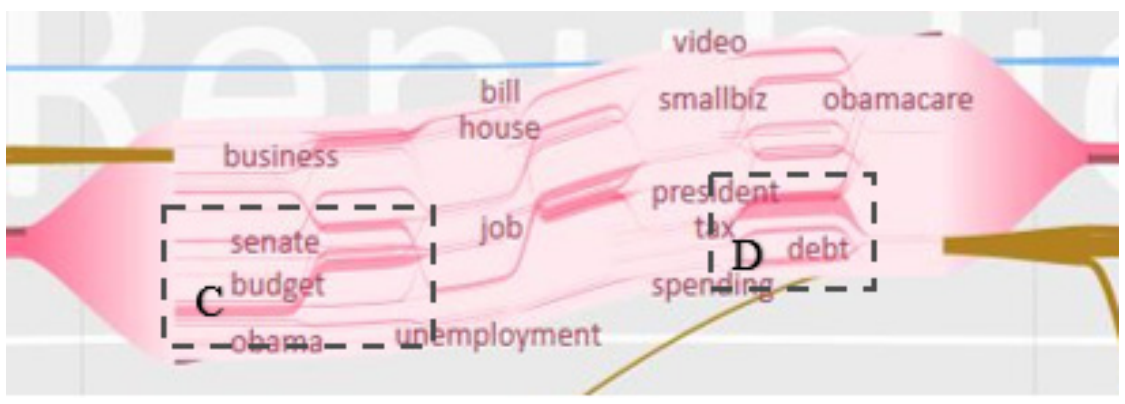

b. 有约束算法效果图

图 6 基于约束的 *t*-SNE 算法效果图

### 4.2 案例分析

为了验证可视分析系统的有效性, 邀请专家使用系统进行分析任务, 观察主题流动的规律, 分析其发生变化的原因, 从更深层次理解主题的流动情况. 通过专家的反馈, 可以对本文的系统进行评估, 在验证系统有效性的同时, 探索改进的方向. 案例中使用的是推特数据集; 在实验中, 用户共分为 4 种: 共和党的众议院成员和参议院成员, 民主党的众议院成员和参议院成员. 每个用户(即推特账号)的信息包括: 发布的消息数量、关注的人和关注它的人. 而每条消息的信息包括: 消息内容、发布用户、发布时间和转发次数等.

#### 4.2.1 案例概览

图 2 所示为美国政党数据集下的系统界面, 可以看到图 2 中 D 区域气泡图中包含 4 个大主题: 文化节日(蓝色), 总统候选人(紫色), 政治政党(黄色)和经济(红色). 结合流视图中的主题流动情况可以发现, 共和党内部政治和经济主题间的流动非常频繁, 可能是经济方面的政策出台导致的现象; 而在民主党和共和党之间的流动的主题类型按照时间依次是经济、政治和文化节日; 此外, 总统候选人主题(紫色)的比例一直比较小, 也没有出现明显的流动现象. 结合以上认知, 专家从流视图开始分析主题的流动情况.

#### 4.2.2 不同主题间的领先-滞后关系分析

专家 *E* 在图 2 右上方的流视图中观察到主题的流动. 以其中 A 区域为例, 在这段时间中, 经济主题和政治政党主题出现了领先-滞后关系, 因此点击展开图 2 中 A 区域的一条主题条带, 展开结果如图 7 所示. 在展开后的图中, 专家 *E* 发现在该时间段里主题中的概念共出现了 3 次聚集现象, 如图 7 中的区域 A, B 和 C. 最先出现的 A 区域中主要包含概念是“pass”, “budget”和“days”等, 接着出现的 B 区域中主要包含“passed”, “obama”和“nobudgetnopay”等概念, 最后出现的 C 区域中则包含“plan”, “house”和“balance”等概念. 根据这 3 个依次出现的聚集概念群, 专家 *E* 进行猜测, 主题从经济流向政治的原因应该与政府的财政预算有关, 应该是经济主题中的一些推特与财务预算相关, 进而引发了关于政府预算的讨论, 使得主题由经济领域流向了政治领域. 为了验证这一猜想, 专家 *E* 点击图 7 中区域 A, B 和 C 的概念, 查看包含这些概念的推特内容.

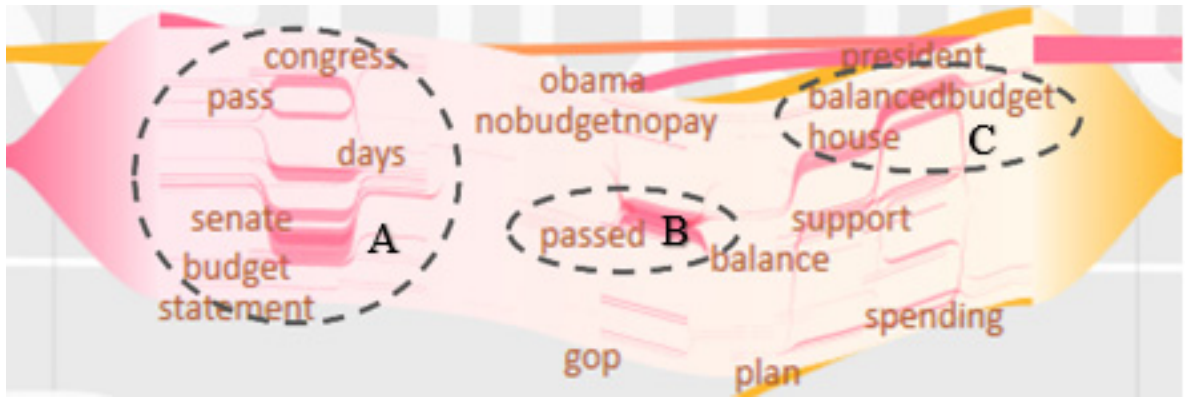

图 7　同文本源主题流展开图

专家 $E$ 从包含图 7 中 A 区域的概念“budget”和“days”的文本中发现，民众们在讨论参议院已经接近 4 年没有公布政府的财政预算；接着，专家 $E$ 从包含图 7 中 B 区域的概念“passed”的文本中发现，民众们仍然在讨论政府没有公布财政预算的问题，但同时发起了一个新的话题“No Budget No Pay”；图 7 中 C 区域出现了一些新的概念，“house”，“balance”等，专家 $E$ 猜测应该是政府对民众的不满给出了回应，并且制定了相应的应对计划，因此查看了包含“house”，“balance”等概念的推特，其真实内容符合猜测：瑞安主席做出了回应，公布了白宫关于平衡预算的计划，引起积极讨论和大量转发.

4.2.3　不同文本源间的主题领先-滞后关系分析

图 2 中 C 区域的 2 个方块代表共和党和民主党 2 个文本源，它们之间用一条细长白色条带分割开. 专家 $E$ 发现文化节日主题很少会在共和党和民主党 2 个文本源之间流动，但是在图 2 右上方流视图的 B 区域，即在 2016 年初，文化节日主题在 2 个政党之间出现了领先-滞后关系，因此专家 $E$ 展开图 2 中 B 区域主题流查看其中的概念如图 8a 所示. 专家 $E$ 发现主题流中的主要是如图 8 中 A 区域所示“https”和“watch”等与主题无关的概念，推测产生的原因是推特中包含了大量视频的链接. 专家 $E$ 删除了这 2 个概念后得到了如图 8b 所示新的投影结果，从新的投影结果中可以看到一些聚堆现象，民众讨论的话题可分为 3 类：图 8 中 C 区域所示白宫的最新消息和相关政策，图 8 中 D 区域所示庆祝节日或纪念日以及图 8 中 B 区域所示国会与奥巴马. 但是，专家 $E$ 在图 2 中 F 区域 Tweets 视图查看包含概念的推特内容，推特转发以及用户关注关系后，发现民主党和共和党成员发布的文化节日主题的推特之间并没有明显的领先-滞后关系. 同时专家 $E$ 发现这些推特中包含大量与政治相关的概念，例如“obama”，“congress”和“bill”等. 包含这些概念的推特有明显的领先-滞后关系. 因此专家 $E$ 推测是因为大量无关概念如“https”等的影响，使主题模型将很多与政治相关的推特误判为与节日主题相关，从而导致节日主题中出现领先-滞后关系. 换言之，节日主题中的领先-滞后关系是由被主题模型误判为节日主题相关的政治类推特造成的.

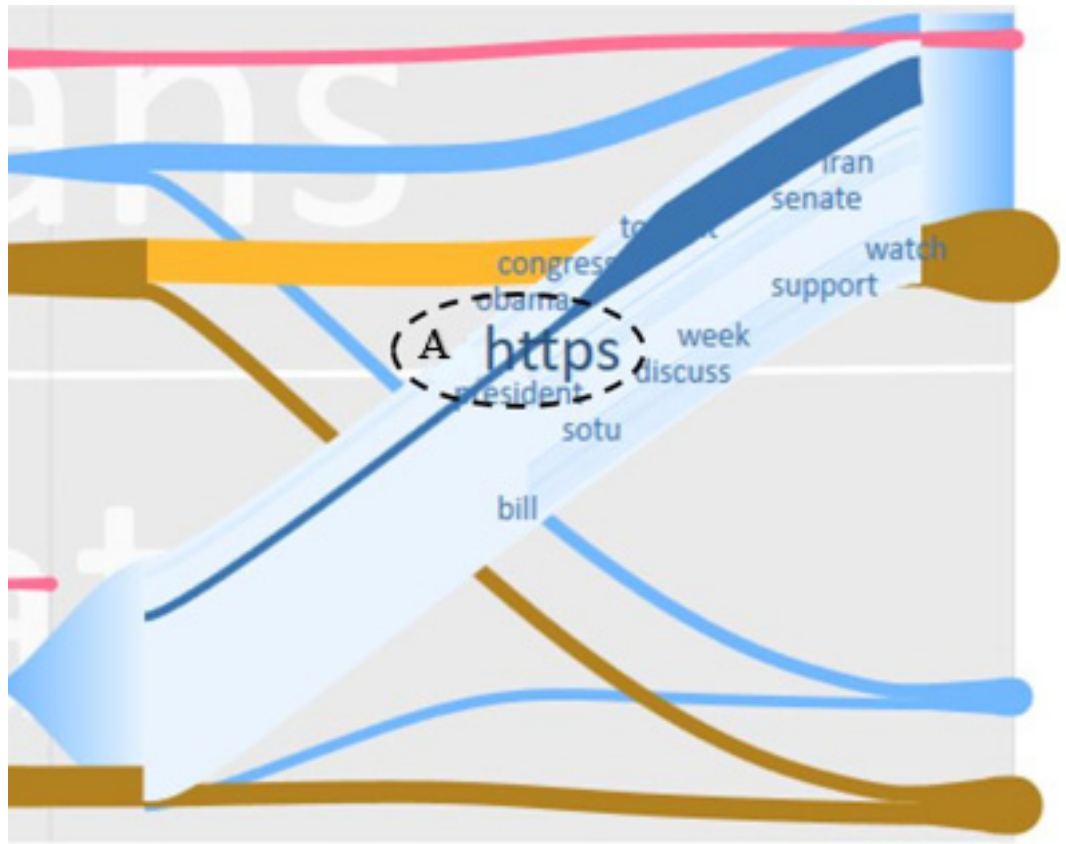

a. 删除无关概念前的主题流展开图

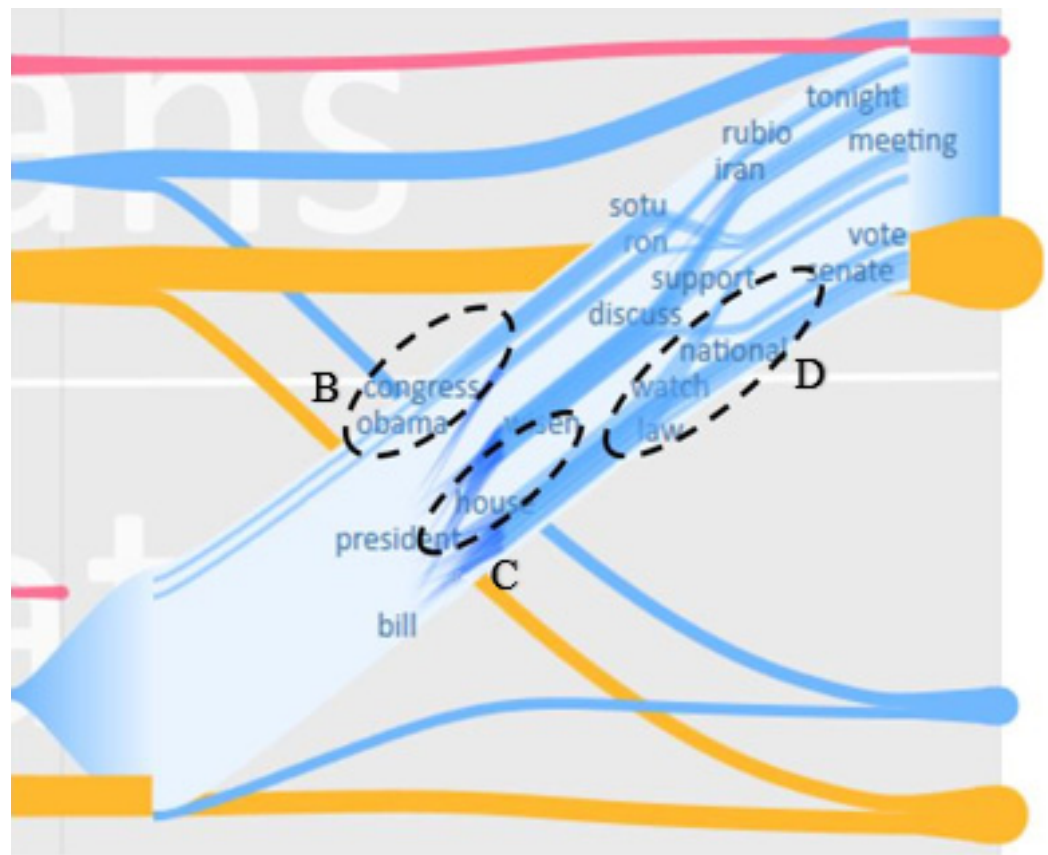

b. 删除无关概念后的主题流展开图

图 8　不同文本源主题流展开图

4.2.4　用户反馈

除了案例分析，本文另请了一位分析者使用了该系统来完成分析工作并给出反馈. 该分析者认可了可视分析系统在帮助分析主题流动方面的能力，特别提到了概念流视图能够帮助他方便快捷地发现数据中主题变化的根本原因. 他给出评论：“通过概念图的帮助，我可以直观地看到主题中蕴含的关键词的变化趋势，还能够轻易地找到并查看相关的文本内容，而不用大量浏览无用的文本数据.”

## 5　总结与未来工作

本文实现了一个交互式的可视分析系统，在 IdeaFlow[31]的基础上，通过概念流可视化方法展现主题中包含的重要概念以及概念分布随时间的变化趋势，帮助专家快速了解主题中的内容. 为了

更好地展现主题中的概念，本文提出了基于异常检测的时间切片算法来定位概念分布发生变化的重要时间段；使用基于约束的 *t*-SNE 降维算法，保证了相邻时间片段的分布稳定性和整个概念流的连贯性．通过领域专家的定性评估和案例研究验证了本文可视分析系统的准确性和有效性．根据专家的反馈，本文还有如下 2 个方面可以改进，首先，利用用户在可视化系统中对主题和主题间相关关系的修改对模型进行更新，从而更准确地提取主题的流动．其次，在分析不同文本源间的主题领先-滞后关系时，如“http”，“image”，“token”之类大量无实际意义的概念由于出现频率高而被提取为重要概念．这些概念导致 2 个语义不相关的主题被计算出较高的相关性．去除这些概念能够改进主题挖掘模型，但不同数据源中包含的无意义概念不同，难以通过简单的停用词方法进行去除．因此，支持用户以半自动的方式去除无意义概念，并迭代式地改进主题挖掘模型是一个值得研究的方向．

## 参考文献(References):